\documentclass[aps,twocolumn,showpacs,preprintnumbers,prl]{revtex4}
\usepackage{graphicx}
\usepackage{amssymb}

\def\TF{^{\rm TF}}
\def\ETF{^{\rm ETF}}
\def\n{n}

\newcommand{\be}{\begin{equation}}
\newcommand{\ee}{\end{equation}}
\newcommand{\bea}{\begin{eqnarray}}
\newcommand{\eea}{\end{eqnarray}}
\newcommand{\bean}{\begin{eqnarray*}}
\newcommand{\eean}{\end{eqnarray*}}

\begin{document}

\title{Ionization potentials in the limit of large atomic number}
\author{Lucian A. Constantin$^1$, John C. Snyder$^1$, John P. Perdew$^2$, and Kieron Burke$^1$}
\affiliation{
$^1$Department of Chemistry, University of California, Irvine, California 92697-2025, USA\\
$^2$Department of Physics and
Quantum Theory Group, Tulane University, New Orleans, LA 70118}
\date{\today}

\begin{abstract}
By extrapolating the energies of non-relativistic atoms and their ions
with up to 3000 electrons 
within Kohn-Sham density
functional theory, we find that the 
ionization potential remains finite and increases across a row, even
as $Z\rightarrow\infty$.  The local density approximation becomes
chemically accurate (and possibly exact) in some cases.
Extended Thomas-Fermi theory matches the shell-average of both the ionization
potential and density change.  Exact results are given in the limit of
weak electron-electron repulsion.
\end{abstract}


\maketitle

A central problem of electronic structure is the calculation of the
ground-state energy of the electrons of any atom, molecule, or solid,
within the non-relativistic Born-Oppenheimer limit.  Density functional
theory (DFT) is a popular choice, balancing computational efficiency
with useful accuracy.
The original DFT was that of Thomas \cite{Thomas} and Fermi \cite{Fermi}, TF theory,
in which a local density approximation is made for the the kinetic energy and
the electron-electron repulsion is approximated by the simple Coulomb energy of
the charge density.
In the 1970's, Lieb and 
co-workers \cite{Lieb1} showed
that the TF
energy 
becomes relatively exact for neutral matter as $Z\rightarrow\infty$ in a specific way.
The energy, $E$, grows in magnitude as $Z^{7/3}$, where $Z$ is the total charge.
For atoms and their ions, the leading corrections in powers of $Z^{-1/3}$
were found by Scott \cite{S52}, Dirac \cite{Dir}, Schwinger and others \cite{Sa80,ESa84}, 
as summarized
by Englert \cite{En88}.
These corrections are given exactly by extended Thomas-Fermi (ETF) theory, which includes
both
the gradient correction for the kinetic energy (one-ninth the von Weisacker functional \cite{vW}),
and the local density
approximation for exchange (LSDX \cite{KS}).

However, TF theory and its extensions
are insufficiently accurate to predict chemical properties\cite{Teller}.
Modern DFT uses the Kohn-Sham (KS) scheme, in which only 
a very small fraction of the total energy, the exchange-correlation (XC), needs be approximated.
But the idea of asymptotic correctness was recently extended to KS,
relating the success of exchange GGA's such as PBE \cite{PBE} for total energies 
to their recovery of the $(Z^1)$ term in the expansion
of the exchange energy \cite{PCSB,EB09}.
The relation between semiclassical
and local density approximations \cite{ELCB} contributed to the creation of PBEsol \cite{PRCVSCZB},
a PBE-like functional that is nearly optimum for solids near equilibrium but not for
atoms and molecules, and to revTPSS \cite{PRCCS}, a nearly optimal semi-local functional
for all three kinds of systems. 

But total electronic energies are irrelevant to chemistry.
Only differences matter, such as the ionization potential of
an atom ($I$ is the energy difference between the positive ion and the neutral)
or the dissociation energy of a chemical bond.
How relevant are asymptotic expansions for these quantities?
The asymptotic expansion for $E$ is in powers of $Z^{-1/3}$, so
if $I$ remains finite as $Z\to\infty$,
the neutral and ion energies must agree for the first {\em seven}
powers in such an expansion, 
a truly remarkable 
balancing act between quantum effects, the Pauli principle, and the
Coulomb forces of nuclear attraction and inter-electron repulsion.
In this letter, we demonstrate by both calculation and analysis that
(i) $I$  has {\em no} limit as $Z\to\infty$, but remains column-dependent
(ii) that each column has
a finite limit;
(iii) the local (spin) density approximation (LSD \cite{KS}) of KS theory
becomes very accurate (if not exact) for $I$ for certain cases;
(iv) ETF theory becomes very accurate (if not exact)
for the average of $I$ over an entire shell;
(v) the shell-averaged difference in density between the neutral and its ion approches
that of TF.  We demonstrate
these statements in the limit of weak interelectron
repulsion.

\begin{figure}
\includegraphics[width=\columnwidth]{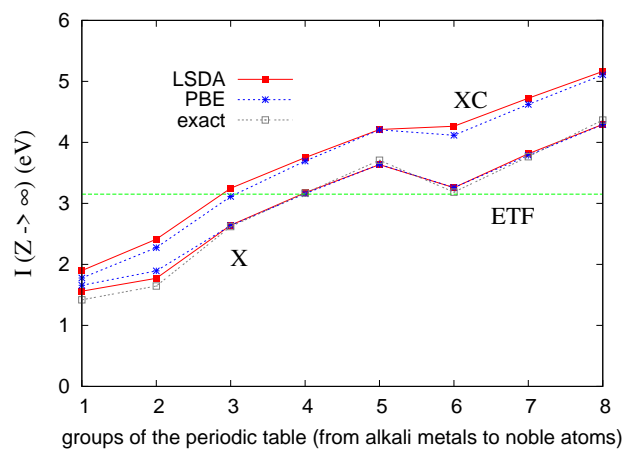}
\caption{Ionization potentials of the main groups in the limit of
large row number of the periodic table,
calculated using exact exchange, the local (spin) density
approximation (LSD), and PBE; ETF denotes extended Thomas-Fermi theory.}
\label{f3}
\end{figure}
Our most important results are shown
in Fig. \ref{f3}.
We plot $I$ from various calculations, extrapolated to infinite
row number, versus 
the column number for main-group elements ($s$ and $p$
valence shells).  
We calculate exchange exactly \cite{KLI,TS}, using the optimized effective potential (OEP, which here should
be indistinguishable from Hartree-Fock \cite{primer}), extrapolating all values to $Z\to\infty$.
At the exchange level, LSD and PBE
are almost exact 
for $p$-valence elements, and are highly accurate but inexact for the $s$-valence cases. 
Furthermore, ETF
yields \cite{En88} a single number (3.15 eV),
very close to the $s$- and $p$-average (3.02 eV).
When correlation is included, gradient effects are slight, and
it is in the regime of large
electron number that approximate density functionals work best, sometimes exactly \cite{PCSB,ELCB,x}.
We speculate that LSDX on accurate densities becomes almost exact in this limit for $p$-shell cases,
that ETF is exact for some shell-average, and that our XC
results are extremely accurate
and practically impossible to calculate with any other method.

To understand why local functionals become accurate in this limit,
begin with total energies of neutral atoms, whose large-$Z$
expansion is
\begin{equation}
E_Q(Z)=-c^{(0)}_q \; Z^{7/3}+0.5 \; Z^2-c^{(2)}_q \; Z^{5/3}+...
\label{el1}
\end{equation} 
where $E_{Q}(Z)$ is the energy of an atom with atomic number $Z$ and
charge $Q$, and the $c^{(j)}$ are coefficients depending on the degree
of ionization, $q=Q/Z$.
We use atomic units throughout.  The neutral coefficients were derived
via semiclassical analysis by Schwinger and
Englert \cite{Sa80,ESa84}.  The TF energy is exactly $-c^{(0)}_q \; Z^{7/3}$.
The
second term \cite{S52}
comes from the $s$-orbitals at the nucleus and must be treated quantum mechanically.
The third term is derivable in ETF theory \cite{Sa80}, of which 2/11 arises from
the gradient correction to the kinetic energy, and 9/11 from LSDX.
When treated in terms of the
potential, the divergence at the nucleus is avoided \cite{En88}.

The extension of these ideas to $I$ has proven 
more difficult.  
Terms of higher order than those shown in Eq. (1)
oscillate \cite{ESa84} with $Z$, as a precursor to the periodic variation of chemical properties
that is missed by ETF, but well-described in KS DFT.
The oscillations in $I$ dominate over trends with $Z^{-1/3}$.
While numerous studies exist in the literature \cite{CGDPFF}
for fixed electron number $N$ with $Z\to\infty$, 
we are interested in $I(Z) = E_1(Z)-E_0(Z)$
as $Z\to\infty$.
Within TF theory, Lieb proved \cite{Lieb2} that $I$ does not grow with $Z$,
and
by considering $c^{(0)}$ as $q\to 0$,
Englert showed    
$I\TF \to 3\Lambda^{-2/3}/7a \approx
1.29$ eV,
where $\Lambda=32.729416$ is a known constant \cite{En88}, and 
$a=(9\pi^2/128)^{1/3}$.  
Even this simple result requires explanation, because
$\mu$, the chemical potential, is zero for the neutral atom in TF theory,
suggesting $I$ should be too.  But the TF energy
is the smooth envelope of $E_{Q}(Z)$ as a function of $q$, whereas the
true energy consists of line segments between integer values \cite{PPLB}.
Thus $\mu=\partial E/\partial q = -I$ for the
exact system, but the TF energy behaves as $q^{7/3}$ for small $q$.
So $\mu_{TF}=0$, but
the better value of 
$I^{TF}$ is 
the energy difference \cite{En88}
with $Q=1$.

\begin{figure}
\includegraphics[width=\columnwidth]{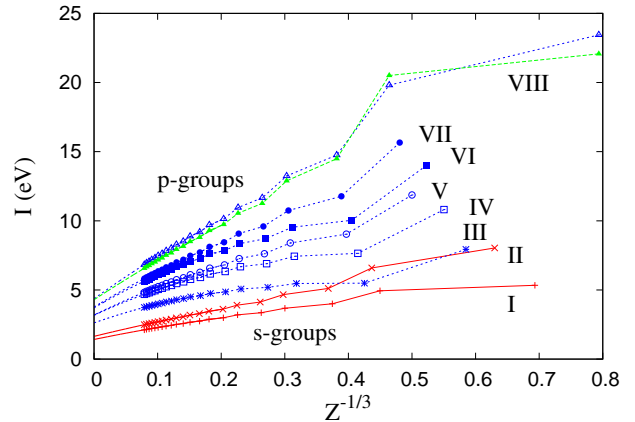}
\caption{OEP ionization potential $I$ (in eV) versus $Z^{-1/3}$ for 
main groups 
of the periodic table. Also shown with green lines is the noble atoms 
LSDX curve.}
\label{f2}
\end{figure}
%
Next we discuss KS DFT, in which the (non-interacting) kinetic energy is not
approximated, but is found
exactly from the KS orbitals.
We perform KS self-consistent calculations for atoms and ions up to 2938 electrons using
LSD and PBE XC functional approximations, as well as the exact 
OEP exchange.
These were done using the Engel code \cite{primer}, but with
tightened convergence criteria and maximum numbers of orbitals, and
a logarithmic radial grid
with 800 points.
In Fig. \ref{f2} we show $I$ versus $Z^{-1/3}$ for each main-group column
of the periodic table. In all cases, the behavior is almost linear
as a function of $Z^{-1/3}$ for all $Z\gtrsim 169$, so we
extrapolated these curves using a parabolic
fit in $Z^{-1/3}$ and 
found the ionization energy for $Z\rightarrow\infty$ as shown
in Fig. \ref{f3}. 
The spherical approximations 
of the density (LSD, PBE) and of the potential (OEP),
used in the Engel code (see  Ref. \cite{TS}), 
give errors less that 0.1 eV for $I$.
We use electronic configurations based 
on the aufbau principle and Madelung rule \cite{Made}. 
For the noble gases,
$Z={n}(n^2+6n+14)/6-\Delta (n)(n/2+1)$,
where $n$ is the row number and $\Delta (n)=0$ for even and 
1 for odd rows.

\begin{table}[htbp]
\footnotesize
\begin{tabular}{|c|ccc|cc|c|c|}
\hline
&\multicolumn{3}{|c|}{$X$}
&\multicolumn{2}{|c|}{$XC$}
&\multicolumn{2}{|c|}{$$}\\
\hline
%
group & LSD & PBE & OEP & LSD & PBE & A & $\langle r \rangle$ \\ \hline  
I & 1.56 & 1.66 & 1.42 & 1.90 & 1.77 & -0.15 & 14.13 \\  
II  & 1.77 & 1.89 & 1.65 & 2.41 & 2.27 & - & 13.56 \\  
\hline
$s$- m.a.d. & 0.13 &  0.24 & 0 & 0.13 & 0 & & \\  
\hline
$s$- avg & 1.67	& 1.78 & 1.54 &	2.16 & 2.02 & -0.15  & 13.85	\\ \hline
III & 2.64 & 2.64 & 2.62 & 3.25 & 3.11 & 0.43 & 10.16 \\  
IV & 3.17 & 3.16 & 3.17 & 3.75 & 3.69 & 0.92 & 9.82 \\  
V & 3.64 & 3.64 &  3.71 & 4.21 & 4.21 & 1.34 & 9.49 \\  
VI & 3.26 & 3.26 & 3.18 & 4.26 & 4.12 & 1.21 & 9.35 \\  
VII & 3.81 & 3.79 & 3.76 & 4.72 & 4.62  & 1.62 & 9.07 \\ 
VIII & 4.29 & 4.29 & 4.37 & 5.16 & 5.11 & - & 8.82 \\  \hline
$p$- m.a.d. & 0.05 & 0.05 & 0 & 0.08 & 0 & & \\  \hline
$p$- avg & 3.47	& 3.46 & 3.47 &	4.23 & 4.14 & 1.10 & 9.45 \\  \hline
\end{tabular}
\label{table2}
\caption{Extrapolated ionization potentials $I$ (eV) of main group elements.
Mean absolute differences (m.a.d.) are taken relative to OEP for X, and PBE for XC.
The last two columns show the electron affinity $A$ (eV) 
(estimated as $I-1/\langle r \rangle$ in atomic units)
and the average radius $ \langle r \rangle$ (bohr)
of the ionization density, in the $Z\rightarrow\infty$ limit, using PBE.
For ETF, $I$=3.15 eV, $ \langle r \rangle=5.6$ \AA, and $A$ = 0.58 eV.
}
\end{table}
To understand in detail the results shown in Fig. \ref{f3}, which are also
tabulated in Table I, we begin
at the exchange level.  Both PBE and LSD exchange are almost
identical to the OEP values for the $p$-group elements, with a maximum difference
between them of 0.02 eV, and of either from OEP of 0.08 eV.  
This is not so for the alkalis and alkali earths, presumably because they
have only one or two electrons {\em outside} a closed shell,
with accompanying self-interaction error of approximate functionals.
The ionization of $p$-elements involves removing electrons from a full
(or almost full) shell with $\sim (n+2)^2/2$ electrons, where $n$ is the row number
for even $n$, and rows $n$ and $n+1$ have the same structure.

\begin{figure}
\includegraphics[width=\columnwidth]{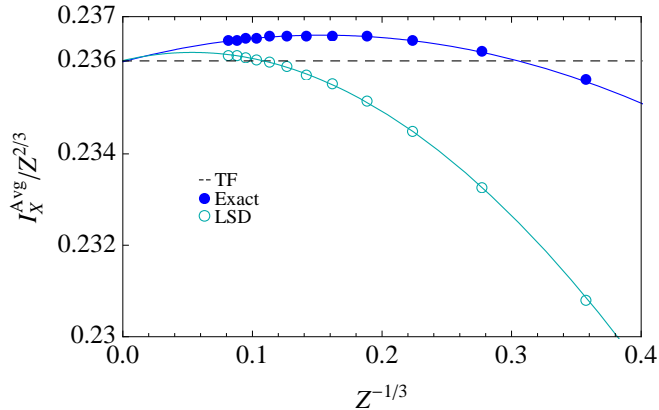}
\caption{Exchange contribution, averaged over shell, to ionization potential for Bohr
atom with many electrons; blue circles are exact, open circles are LSDX on exact density,
and black dashed line is LSDX on TF density.  Solid lines are cubic fits to the
last ten circles.}
\label{fJohn}
\end{figure}
In fact, 
Englert also showed that the TF result is {\em not} correct as $Z\to\infty$.
The terms of $O(Z^{5/3})$ in Eq. (1) also yield a finite contribution, which 
is included in ETF,
yielding an
$I$ of 3.15 eV, very close to the average over both $s$ and $p$-shell values (3.02 eV).

To check this is no accident, consider the simpler system of atoms
with an infinitesimal electron-electron repulsion, $\lambda$, sometimes called Bohr atoms.
The orbitals are hydrogenic, requiring no self-consistency and simplifying the integrals \cite{HL}.
One finds that $I\TF$ is exact for large $Z$ at $\lambda = 0$.
In Fig. \ref{fJohn}, we show the exchange correction (divided by $\lambda$)
to $I$ for LSDX applied to the TF density (yielding
$8 (2/3)^{1/3}/(3\pi^2) \approx 0.2360$), to 
the exact densities (each averaged over entire shells), and exactly.
All three match as $Z\to\infty$, but
a small error remains if, e.g., just the $s$-shell is used.
Thus we
speculate that, for real atoms,
LSDX (in a KS calculation) 
matches the average over the {\em entire} shell as $Z\to\infty$.

Next, we discuss the DFT calculations with correlation, which
remains finite as $Z\to\infty$ and varies across a row.
The differences between PBE and LSD are relatively small, giving greater
confidence in both.  The maximum deviation between them for $p$-elements
is 0.14 eV, comparable to the deviations of these functionals
at the exchange-only level from OEP
for the alkali and alkali earths.  Thus
the gradient corrections are {\em not} vanishing, suggesting
that while both calculations are accurate, neither is exact.  The PBE average,
3.61 eV,
is our best estimate of a universal ionization potential, defined
as the limit of $I$ averaged over the $n$-th shell, as $n\to\infty$.

The other major descriptor of chemistry is
the electron affinity $A(Z) = E_0(Z)-E_{-1}(Z)$.  Within
LSD or PBE, the first negative atomic ion of energy $E_{-1}(Z)$ has no
stable solution, but $A(Z)$ can still be estimated 
\cite{P88}
via a charged conductor model, in which $I-A=1/\langle r \rangle$,
and $\langle r \rangle$ is the centroid of the added charge.
Define the radial ionization density as
\be
\Delta \n_R(Z,r)=4\pi r^2\,\left( \n_0(Z,r) - \n_1(Z,r)\right),
\ee
which integrates to $1$.  
Then choose
$\langle r\rangle=\int^{\infty}_0 \; dr\;r  \Delta n_R(Z,r)$.
Table I shows PBE $Z\rightarrow\infty$ limits for $I$, $\langle r\rangle$, and
$A$.  Averaging over $s$ and $p$, our
best estimate for a universal value of $A$ is 0.78 eV.

\begin{figure}
\includegraphics[width=\columnwidth]{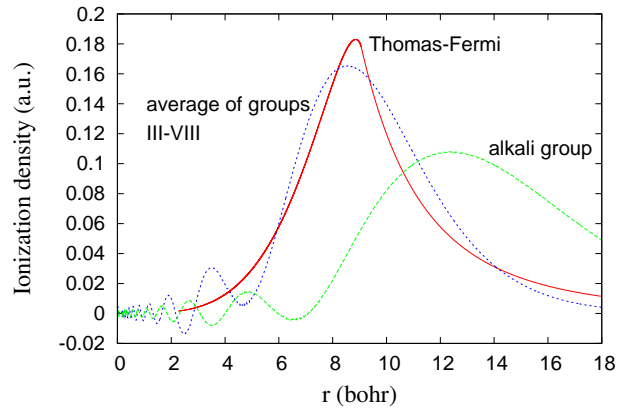}
\caption{ $q=1$ ionization density, $4\pi r^2 (n_{atom}(r)-n_{ion}(r))$, as $Z\to\infty$ for the
average
over the groups III-VIII, for the alkali series, and within TF theory. We use LSDX KS densities.
}
\label{f6}
\end{figure}

We next extrapolate the ionization density via
\begin{equation}
\Delta n_R(Z,r)\approx \beta \Delta n_R(Z_0,\beta r)+\gamma d[\Delta n_R(Z_0,r)]/dr,
\label{ew1}
\end{equation}
which correctly integrates over $r$ to $1$.
Here $Z_0=2935$, $\beta=1+b (Z^{-1/3}-Z^{-1/3}_0)$, and
$\gamma=c(Z^{-1/3}-Z^{-1/3}_0)$, with fit parameters 
$b=5$, and $c=-2$. 
Finally, we also averaged over the 6 $p$-shell
curves, to find the results shown in Fig. \ref{f6}.
The TF solution for the infinitesimally charged ion has a finite size\cite{En88,Lieb2}, i.e.,
\begin{equation}
r_c=\lim_{Z\to\infty} r_0(Z)=a\, \Lambda^{2/3} \approx 9.0588 \; \rm{bohr} \approx 4.8\; \rm{\AA}.
\end{equation}
Beyond this radius, $\Delta n_R\TF(r)$ is just the radial density of
the neutral, which has reached its asymptotic form, decaying as $1/r^4$.
The maximum of this curve is about 0.1830 at $r=8.855 \; \rm{bohr}$.
The agreement between the extrapolated $p$-shell densities and the TF theory is remarkably good,
but not exact,
while the extrapolated alkali ionization density is very different.
We speculate that averaging over an entire shell would yield perfect agreement,
as we find numerically for the Bohr atom.

\begin{figure}
\includegraphics[width=\columnwidth]{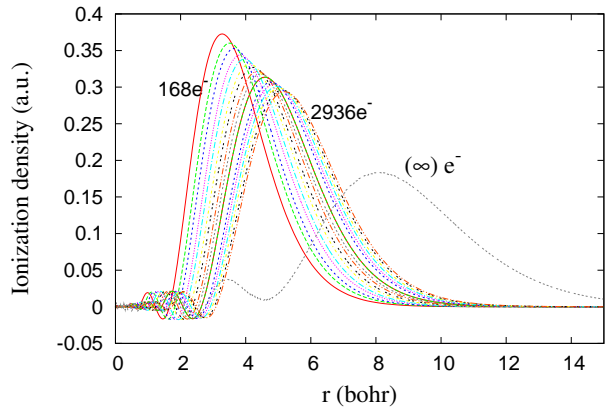}
\caption{Same as Fig. \ref{f6}, but for the noble-gas column of the periodic table
at various finite $Z$ and in the limit $Z\rightarrow\infty$.}
\label{f4}
\end{figure}
Finally, we justify why such large 
atomic numbers
(larger by a factor of 10 than those of Ref. \cite{CMA}) are needed
to get these results.
Because of the scaling with $Z^{-1/3}$, even $Z=125$ only makes $Z^{-1/3}=0.2$, while
$Z > 1000$ brings $Z^{-1/3}$ below 0.1, making the extrapolation
much more reliable.  
In Fig. \ref{f4},
we show accurate ionization densities
for the 8-th column of the extended
table at finite $Z$.
The scaling of the
TF ionization density is quite different from that of the exact solutions:
Before extrapolation, even at $Z=2935$, the TF ionization density agrees
much better with that of the {\em alkalis}, not the $p$-shell average.
For the same reasons, having HF energies for only $Z \lesssim 100$, Englert 
erroneously concluded that $I\ETF$ was the limit of the alkalis, not the shell-average
(see Fig. 4-8 and its discussion of Ref. \cite{En88}).

        Thomas-Fermi theory produces the first term of Eq. (1) and extended TF yields
an average
$Z\to\infty$ limit for the ionization energy, but no periodic variation
of chemical properties and no binding \cite{Teller} of atoms to form molecules or solids.
Within non-relativistic KS theory, any reasonable approximation to the XC energy with
the correct uniform-density limit for exchange will produce the total-energy expansion
of Eq. (1) and a finite column-dependent $Z\to\infty$ limit for the ionization energy.
It appears that LSD is extremely accurate and possibly
exact in certain cases for $I$.  We have shown this for shell-averages in the limit of weak inter-electron
repulsion.  But in that case, the last shell is spread throughout the entire atom, and
average gradients contributing to ionization vanish as $Z\to\infty$, which is not true for
real atoms.

Thus we have established that,
in the large-$Z$ limit, the periodic table becomes \emph{perfectly} periodic. 
Moreover, local approximations appear to become exact, even for energy differences
that are relatively vanishingly small in this limit.  These are
new, numerically relevant, exact conditions
that approximate functionals should satisfy.

        All conclusions are based upon numerical calculations and extrapolation.
Proving them rigorously is a challenge to mathematical physics.
This work was supported by DOE grant DE-FG02-08ER46496 at Irvine, and
NSF (Grants DMR-0501588 and DMR-0854769) at Tulane.

\end{document}